\documentclass[letterpaper, 10 pt, conference]{ieeeconf}  

\IEEEoverridecommandlockouts                              

\usepackage{graphics}
\usepackage{color}

\usepackage{hyperref}
\usepackage{algorithm}
\usepackage{algpseudocode}
\usepackage{mathtools}
\DeclarePairedDelimiter{\ceil}{\lceil}{\rceil}

\DeclarePairedDelimiter{\round}{\lceil}{\rfloor}

\DeclareMathOperator*{\argmax}{arg\,max}
\DeclareMathOperator*{\argmin}{arg\,min}

\usepackage{amsthm}
\theoremstyle{plain}
\newtheorem{theorem}{Theorem}
\theoremstyle{definition}
\newtheorem{defn}{Definition}
\theoremstyle{remark}
\newtheorem{rem}{Remark}
\usepackage{mathptmx} 
\usepackage{amsmath} 
\usepackage{amssymb}  
\usepackage[margin=0.79in]{geometry}

\title{\LARGE \bf
Compute-and-Forward: Finding the Best Equation
}

\author{Saeid Sahraei and Michael Gastpar$^{1}$
\thanks{$^{1}$Both authors are with the Department of Computer and Communication Sciences,
EPFL, Switzerland
        {\tt\small saeid.sahraei@epfl.ch}$\;\;$ , $ \;\;$
        {\tt\small michael.gastpar@epfl.ch}}%
}

\begin{document}

\maketitle
\thispagestyle{empty}
\pagestyle{empty}

\begin{abstract}

Compute-and-Forward is an emerging technique to deal with interference.
It allows the receiver to decode a suitably chosen integer linear combination
of the transmitted messages. The integer coefficients should be adapted
to the channel fading state. Optimizing these coefficients is a Shortest Lattice
Vector (SLV) problem. In general, the SLV problem is known to be prohibitively
complex. In this paper, we show that the particular SLV instance resulting from
the Compute-and-Forward problem can be solved in low polynomial complexity and
give an explicit deterministic algorithm that is guaranteed to find the optimal
solution. 

\end{abstract}

\section{INTRODUCTION}

Compute-and-Forward is an emerging paradigm for dealing with interference in wireless multiuser channels. Contrary to previous approaches, Compute-and-Forward does not view interference as inherently undesirable. The key idea is that the relay nodes aim at recovering integer linear combinations of transmitted codewords instead of attempting to decode the individual codewords and treating interference as noise. Nested lattice codes ensure that an integer linear combination of codewords is a codeword itself. This method is shown to considerably enhance the achievable rates compared to other techniques \cite{nazer2011compute,nam2008capacity,wilson2010joint}.

\begin{figure}[h]
\includegraphics[width=8cm]{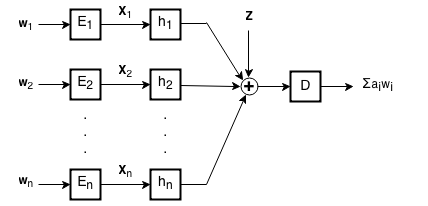}
\caption{$n$ transmitters send their signals to one relay. The relay decodes an integer linear combination of the codewords.}
\label{fig:fig1}
\end{figure}

In a typical Compute-and-Forward scenario, ${n}$ transmitting nodes, each with transmission power $P$, share a wireless channel to send their messages to a relay node. The relay receives a noisy linear combination of the transmitted messages, namely
$${y} = \sum_{i=1}^nh_ix_i + z$$
where $y$ is the received signal, $x_i$ and $h_i$ respectively represent the signal transmitted by node $i$ and the effect of channel from this node to the relay and $z$ denotes additive white Gaussian noise of unit variance. The receiver then recovers an integer linear combination of the transmitted codewords (\autoref{fig:fig1}).
It is proved in \cite{nazer2011compute} that the achievable rate for the Compute-and-Forward scheme is at least as large as:
\begin{equation}
R({\bf h}, {\bf a}) =  \frac{1}{2}\log^+ \left(\left(\|{\bf a}\|^2 - \frac{P|{\bf h}^T{\bf a}|^2}{1+P\|{\bf h}\|^2} \right)^{-1} \right)
\label{eqn:CF}
\end{equation}
where ${\bf a}$ describes the coefficient vector of the linear combination of messages to be recovered at the relay node.

From the perspective of a single receiver, a reasonable choice for the vector ${\bf a}$ is one that maximizes the achievable rate given by (\ref{eqn:CF}), that is:
\begin{equation}
{\bf a}^* = \argmax_{{\bf a}\in \mathbb{Z}^n\backslash\{\bf 0\}}  \frac{1}{2}\log^+ \left(\left(\|{\bf a}\|^2 - \frac{P|{\bf h}^T{\bf a}|^2}{1+P\|{\bf h}\|^2} \right)^{-1} \right).
\label{eqn:CFopt}
\end{equation}

It is shown in \cite{nazer2011compute} that ${\bf a}^*$ satisfies 
\begin{equation}
\|{\bf a^*}\| \le \sqrt{P\|{\bf h}\|^2+1}.
\label{eqn:CFbound}
\end{equation}
Given the bounded interval, one can attempt to solve problem (\ref{eqn:CFbound}) by exhaustive search. For a small number of users $n$ this is feasible, but it quickly becomes prohibitively complex as $n$ grows large. 

With slight manipulation, this optimization problem can be rewritten as 
\begin{equation}
{\bf a}^* = \argmin_{{\bf a}\in \mathbb{Z}^n\backslash\{\bf 0\}}  f({\bf a})={\bf a}^T {\bf G a}
\label{eqn:CFmain}
\end{equation}
where
\begin{equation*}
{\bf G}  =  (1+P\|{\bf h}\|^2){\bf I} - P{\bf h}{\bf h}^T.
\end{equation*}

is a positive definite matrix. This is an instance of a well-known problem in discrete mathematics called the Shortest Lattice Vector problem. The SLV problem is known to be NP-hard in its general form, that is for a general positive definite matrix ${\bf G}$. Due to this fact, it has been suggested that approximation algorithms must be used in order to solve (\ref{eqn:CFmain}). The problem with such algorithms is that the best known polynomial complexity approximation algorithms for the SLV problem have exponential approximation factors. The most famous among them are the celebrated LLL algorithm \cite{lenstra1982factoring} and its extensions, most notably \cite{gama2008finding}. In fact, Khot in \cite{khot2004hardness} has shown that assuming $NP\nsubseteq RP$ , no constant factor approximation algorithm can be found for the SLV problem which runs in polynomial complexity. Other results on hardness of the SLV problem have been found, for instance in \cite{dadush2011enumerative,dadush2013algorithms,alekhnovich2005hardness}.

On the bright side, efficient algorithms for special lattices have been known for a long time. For instance Gauss found an algorithm for solving the SLV problem in dimension two. Conway in \cite{bannai1999sphere} provides exact algorithms for a class of root lattices in higher dimensions. Based on \cite{conway1992low} McKilliam \cite{mckilliam2012finding} showed that if an obtuse superbase for a lattice is known, the shortest vector can be found in polynomial complexity.

Since the publication of Compute-and-Forward paper \cite{nazer2011compute}, there has been a few attempts to tackle the special case of SLV problem which appears in (\ref{eqn:CFmain}). Performance of the proposed methods are mostly exhibited through simulation results or heuristic arguments and NP-hardness of the problem in hand is typically the underlying assumption \cite{hejazi2013simplified,richter2012efficient,zhou2014quadratic}.

In this work we show that the special case of SLV problem which appears in Compute-and-Forward admits an exact solution of polynomial complexity. 
The following theorem, albeit provable mostly by elementary manipulations of integer inequalities, establishes an important fact that provides the foundation of our SLV algorithm. (Note that the operator $\round{\cdot}$ rounds each element of its vector input to the closest integer.)

\begin{theorem}
{\label {THM1}}
The solution to (\ref{eqn:CFmain}) satisfies
$${\bf a^*} - \frac{1}{2}\mathbf{1}< {\bf h}x < {\bf a^*} + \frac{1}{2}\mathbf{1}$$
and thus 
$${\bf a^*} = \round{{\bf h}x}$$
for some $x \in \mathbf{R}$. Or ${\bf a^*}$ must be a standard unit vector, up to a sign.\\
\end{theorem}

It follows from \autoref{THM1} that for the special lattices of interest, the shortest vector can be obtained by solving an optimization problem over only one variable. Equation (\ref{eqn:CFbound}) tells us the search only has to be done over a bounded region. An individual examination of the standard unit vectors must also be performed. This will significantly reduce the number of candidate vectors ${\bf a}$. We will find an upper bound on this number, propose a method to enumerate all such candidates and find the one that minimizes $f$ as defined in (\ref{eqn:CFmain}). We will prove that the complexity of our algorithm is
$$O\left(n^2\sqrt{1+P\|{\bf h}\|^2}\right).$$

\begin{rem}
{\label{Remark1}}
The formula given by {\autoref{THM1}} has some resemblance to the results of \cite{mckilliam2008linear} and \cite{mckilliam2010linear}. However the span of these works are Coxeter lattices and the goal is to find faster algorithms for problems which are already known to be polynomially solvable.
\end{rem}

The rest of the paper is organized as follows: First we define the notation used throughout the paper. In Section {\ref{sec:thealgorithm}} we introduce an algorithm which solves the optimization problem (\ref{eqn:CFmain}) based on {\autoref{THM1}}. The complexity of the algorithm will then be calculated. This will be followed by a generalization of \autoref{THM1} and the corresponding algorithm to the case of MIMO Compute-and-Forward respectively in Sections {\ref{sec:generalization}} and \ref{sec:alg2}. An analysis of the complexity of this algorithm will be given in Section {\ref{sec:complexity}}. In Section {\ref{sec:Proof1}} the proof of {\autoref{THM2}} which is a generalized version of \autoref{THM1} is given. Finally we will conclude our work in Section {\ref{sec:conclusion}}.

\section{NOTATION}
We use boldface lowercase letters to denote vectors. All vectors are assumed to be vertical. In particular we use $\mathbf{1}$ to denote the all-ones vector and $\mathbf{0}$ for the all-zero vector. Boldface capital letters represent matrices. Scalars are written with plain letters. For example, for a matrix ${\bf A}$ we use $A_{ij}$ to refer to the element in its $i$-th row and $j$-th column. Similarly, for the vector ${\bf a}$, we denote its $i$-th element by ${a}_i$. When referring to indexed vectors, we use boldface letters. For instance, ${\bf a}_{i}$ denotes the $i$-th vectors, whereas ${a}_{ij}$ indicates the $j$-th element of the $i$-th vector. 
\\For an $n\times m$ matrix ${\bf A}$ and for a set ${\pi \subseteq \{1,\dots,n\}}$ we define ${\bf A}_\pi$ as the submatrix of ${\bf A}$ which consists of the rows indexed in ${\pi}$. For a vector ${\bf a}$, we define ${\bf a}_\pi$ in a similar manner. For an $n\times n$ matrix {\bf A} we use $diag({\bf A})$ to denote a vector consisting of its diagonal elements.\\
All the vector inequalities used throughout the paper are elementwise. The operator $\ceil{\cdot}$ returns the smallest integer greater or equal to its input. The two operators $\lceil\cdot\rfloor$ and $\lfloor\cdot\rceil$ return the closest integer to their input. Their difference is at half-integers: the former rounds the half-integers up and the latter rounds them down. We use $\|\cdot\|$ to represent the 2-norm of a vector. Finally, $\mathbb{R}$ represents the set of real numbers and $\mathbb{Z}$ the set of integers.

\section{ALGORITHM I}
\label{sec:thealgorithm}
In this section we provide an efficient algorithm for solving the optimization problem (\ref{eqn:CFmain}). We will show that the algorithm runs in polynomial complexity in $n$. For the sake of convenience we define $\psi =\sqrt{1+P\|{\bf h}\|^2}$.\\

In line with \autoref{THM1} define ${\bf a}({{x}}) = \round{{\bf h}x}$. Note that {\autoref{THM1}} reduces the problem to a one-dimensional optimization task. Since every ${a}_i(x)$ is a piecewise constant function of $x$, so is the objective function $f$. Overall, the goal is to find a set of points which fully represent all the intervals in which $f$ is constant and choose the point that minimizes $f$. Being a piecewise constant function, $f$ can be represented as:
\begin{equation}
  f({\bf a}(x))=\begin{cases}
    f_i\;\; ,& \text{if $\xi_i< x< \xi_{i+1}$ ,  $i = \dots, -1,0,1,\dots$}\\
    h_i\;\; ,& \text{if $x = \xi_i$ ,  $i = \dots, -1,0,1,\dots$}\\
  \end{cases}
  \label{eqn:1}
\end{equation}
$\xi_i$ values are sorted real numbers denoting the points of discontinuity of $f$. Since $f$ is a continuous function of ${\bf a}$, these are in fact the discontinuity points of ${\bf a}(x)$ (or a subset of them) or equivalently the points where $a_i(x)$ is discontinuous, for some $i=1\dots n$. Since we have that ${a_i}(x)= \round{{h}_ix}$, the discontinuity points of $a_i(x)$ are the points where $h_ix$ is a half-integer. Or equivalently the points of the form $x= \frac{c}{|h_i|}$ where $c$ is a half-integer and $h_i\neq 0$. To conclude this argument, we write:

\begin{equation}
\begin{split}
\xi_i &\in \left\{ \frac{c}{|h_j|}\;\middle| \;j=1\dots n \;, h_j \neq 0 \; , \; c-\frac{1}{2} \in\mathbb{Z} \right\}\\
i &= \dots, -1 , 0 , 1 ,\dots
\end{split}
\label{eqn:2}
\end{equation}
We can also see from \autoref{THM1} that ${\bf a}^*$ satisfies ${\bf a^*} - \frac{1}{2}\mathbf{1}<{\bf h}x < {\bf a^*} + \frac{1}{2}\mathbf{1}$ for some $x\in \mathbb{R}$. Hence any $x$ satisfying 
\begin{equation}
{a}_i^* - \frac{1}{2}< xh_i <{a}_i^* + \frac{1}{2}\;\;, \;\;i=1\dots n \;,\;h_i \neq 0
  \label{eqn:3}
\end{equation}
 minimizes $f$. As a result, $x$ belongs to the interior of an interval and not the boundary. Therefore, in the process of minimizing $f$, one can ignore the $h_i$ values in (\ref{eqn:1}), check all $f_i$ values and choose the smallest one.
$$\min_{{\bf a}\in \mathbb{Z}^n\backslash\{{\bf 0}\}} f({\bf a}) = \min_{i = \dots -1,0,1\dots} f_i$$
Since $\frac{\xi_i+\xi_{i+1}}{2}$ belongs to the interval $(\xi_i , \xi_{i+1})$, we can rewrite $f_i$ as $f_i = f({\bf a}(\frac{\xi_i+\xi_{i+1}}{2}))$. Thus:
\begin{equation}
\min_{{\bf a}\in \mathbb{Z}^n\backslash\{{\bf 0}\}} f({\bf a}) = \min_{i = \dots -1,0,1\dots} f({\bf a}(\frac{\xi_i+\xi_{i+1}}{2}))
  \label{eqn:4}
\end{equation}
On the other hand, (\ref{eqn:CFbound}) tells us that we do not need to check the whole range of x. It follows from the constraint $\|{\bf a}\| \le \psi = \sqrt{1+P\|{\bf h}\|^2}$ that $|a_i| \le \psi$ and thus, from (\ref{eqn:3}) we have
\begin{align*}
-\frac{1}{|h_i|}(\psi+ \frac{1}{2})&< x < \frac{1}{|h_i|}(\psi + \frac{1}{2}) \;\;, \;\;i=1\dots n \;,\;h_i \neq 0 \\
\Rightarrow -\beta(\psi+ \frac{1}{2})&< x < \beta(\psi + \frac{1}{2})
\end{align*}
where 
\begin{equation}
\beta = \min_{i=1\dots n \;,\;h_i \neq 0}\frac{1}{|h_i|}
\end{equation}
It follows from this expression and equation (\ref{eqn:2}) that the largest $\xi_{i+1}$ that we need to check in equation (\ref{eqn:4}) is $\beta(\ceil{\psi} + \frac{1}{2})$. Similarly the smallest $\xi_i$ to be checked is $-\beta(\ceil{\psi} + \frac{1}{2})$.
We can now rewrite equation (\ref{eqn:4}) as:
\begin{equation}
\min_{{\bf a}\in \mathbb{Z}^n\backslash\{{\bf 0}\}} f({\bf a}) = \min_{\substack{i = \dots -1,0,1\dots \\ \xi_i\ge-\beta(\ceil{\psi} + \frac{1}{2})\\ \xi_{i+1}\le\beta(\ceil{\psi} + \frac{1}{2})}} f({\bf a}(\frac{\xi_i+\xi_{i+1}}{2}))
\label{eqn:6}
\end{equation}
Using equation (\ref{eqn:2}) we can translate the constraints in equation (\ref{eqn:6}) into:
\begin{align}
 \frac{c}{|h_j|} &\le \beta(\ceil{\psi} + \frac{1}{2})\Rightarrow c \le \ceil{\psi}+\frac{1}{2} \;,\;j=1\dots n\;,\; \text{  and}\\
  \frac{c}{|h_j|} &\ge -\beta(\ceil{\psi} + \frac{1}{2})\Rightarrow c \ge -(\ceil{\psi}+\frac{1}{2})\;,\;j=1\dots n
\end{align}
By defining the sets $\Phi_j$ , $j=1\dots n$ and the set ${\Phi}$ as follows:
\begin{align}
\begin{split}
\Phi_j &=  \left\{ \frac{c}{|h_j|} \;\middle| \; |c| \le\ceil{\psi}+\frac{1}{2}\;,\; c-\frac{1}{2} \in\mathbb{Z} \right\}\\
j&= 1\dots n \; , \; h_j\neq 0
\end{split}\\
\Phi_j &= \emptyset \;\; , \;j= 1\dots n \; , \; h_j = 0\\
\Phi &= \bigcup_{j=1}^n\Phi_j
\end{align}
and after sorting the elements of $\Phi$, we can write the equation (\ref{eqn:6}) as 
\begin{equation}
\min_{{\bf a}\in \mathbb{Z}^n\backslash\{{\bf 0}\}} f({\bf a}) = \min_{\xi_i\;,\;\xi_{i+1}\in\Phi} f({\bf a}(\frac{\xi_i+\xi_{i+1}}{2}))
\label{eqn:9}
\end{equation}
Thus, the algorithm starts by calculating the sets $\Phi_j$ and their union $\Phi$, sorting the elements of $\Phi$ and then running the optimization problem described by equation (\ref{eqn:9}). The standard unit vectors will also be individually checked. The number of elements in $\Phi_j$ is upper-bounded by $2\ceil{\psi}+2$ and thus the number of elements in $\Phi$ is upper-bounded by $n(2\ceil{\psi}+2)$. Also the value of $f$ can be calculated in $O(n)$. So the complexity of the algorithm is $O(n^2\sqrt{1+P\|{\bf h}\|^2})$. The procedure is summarized in Algorithm \ref{Algorithm1}.

\begin{algorithm}
\caption[caption]{Finding the optimal coefficient vector}
\begin{algorithmic}[1]
\Statex {{\bf Input:} Channel vector ${\bf h}$ and transmission power P}
 \Statex {\bf Output: }{$\bf {a^*}$}
\Statex ${\bf \underline {Initialization:}}$
 \State ${\bf u}_i:=$ standard unit vector in the direction of $i$-th axis
 \State $\psi := \sqrt{1+P\|{\bf h}\|^2}$
\State $\Phi=\emptyset$
\State ${\bf G}:=({1+P\|{\bf h}\|^2}){\bf I} - P{\bf hh}^T$
\State $f({\bf a}):={\bf a}^T{\bf G}{\bf a}$
\State $f_{\min} = \min(diag({\bf G}))$
\State ${\bf a^* }= {\bf u}_{\argmin(diag({\bf G}))}$
\Statex{}
\Statex {\bf \underline {Phase 1:}}
\For{all $i \in \{1,\dots ,n\}$, and ${h_i \neq 0}$}
  \For{all ${c}$ , $|{c} |\le \ceil{\psi}+\frac{1}{2}$ and $ {c}-\frac{1}{2} \in\mathbb{Z} $}
\State	calculate ${x}=\frac{c}{|h_i|}$
\State	Set $\Phi = \Phi \cup\{x\}$

\EndFor
\EndFor
\Statex{}
\Statex {\bf \underline {Phase 2:}}
\State sort $\Phi$
\For{every two consecutive members of $\Phi$}
\State	calculate ${x}$ = average of the two points
\State	calculate ${\bf a} = \round{{\bf h}x}$
	\If {$f({\bf a})<f_{\min}$ AND ${\bf a}$ is not the all zero vector}
\State	set ${\bf a^* = a}$
\State	set $f_{\min}= f({\bf a})$

\EndIf
\EndFor
\State \Return {${\bf a^*}$}

\end{algorithmic}
\label{Algorithm1}
\end{algorithm}

\section{A GENERALIZATION TO MIMO COMPUTE-AND-FORWARD}
\label{sec:generalization}

In this section we provide a generalization of \autoref{THM1} and the corresponding algorithm by relaxing several constraints that we imposed on the structure of the Gram matrix. The generalized theorem can be applied to maximize the achievable rate of MIMO Compute-and-Forward studied in \cite{zhan2009mimo}. The scenario is very similar to what mentioned in the introduction, with the difference that the relay node now has multiple antennas. The objective remains the same: decode the best integer linear combination of the received codewords. Assume there are $n$ transmitters with transmission power $P$ and the receiver node has $k$ antennas. Let ${\bf h}_i$ be the channel vector from the transmitting nodes to the $i$-th antenna of the relay. Also, let ${\bf H}$ be the $n\times k$ matrix whose columns are the ${\bf h}_i$ vectors. It directly follows from the results of \cite{zhan2009mimo} that the achievable rate satisfies the following equation:
\begin{equation}
R({\bf a}) = - \frac{1}{2}\log {\bf a}^T{\bf Ga}
\end{equation}
where
\begin{equation}
{\bf G} = {\bf WRW}^T.
\label{eqn:CFMIMO}
\end{equation}
Here ${\bf W \;\;}$ is a unitary matrix in $\mathbb{R}^{n\times n}$ whose columns are the eigenvectors of ${\bf HH}^T$, and ${\bf R}$ is a diagonal square matrix with the first $k$ diagonal elements satisfying
$$r_i = \frac{1}{1+P\gamma_i^2}\;\;,\;\; i=1\cdots k$$
 and the last $n-k$ diagonal elements equal to 1. Finally, $\gamma_i^2$ is the $i$-th eigenvalue of ${\bf HH}^T$ (same order as the columns of ${\bf W}$).\\
 The goal is to maximize the achievable rate or equivalently, to find 
 $${\bf a}^* = \argmin_{{\bf a}\in \mathbb{Z}^n\backslash\{\bf 0\}}  {\bf a}^T {\bf G a}$$
as in the single antenna case.

We first mention a generalization of \autoref{THM1} and next we show that the Gram matrix which appears in this optimization problem satisfies the constrains of the new theorem. 
To begin with, we define the following:
\begin{defn}[$DP^k$ decomposable matrices]
We call a positive definite matrix $DP^k$ decomposable if it can be written as ${\bf G =D - P}$ where ${\bf P}$ is a positive semi-definite matrix of rank $k$ and ${\bf D}$ is a diagonal matrix. We call such a representation a $DP^k$ decomposition of the matrix ${\bf G}$.\\
\end{defn}

It follows from the definition that all diagonal elements of ${\bf D}$ are strictly positive. This is due to the fact that ${\bf G}$ is positive definite and ${\bf P}$ is positive semi-definite. Furthermore, we find it convenient to write ${\bf P}$ as ${\bf P} = {\bf VV^T}$ where ${\bf V}$ is an $n\times k$ matrix whose columns are linearly independent. Such a decomposition is not unique, but our arguments will be valid regardless of how the matrix ${\bf V}$ is chosen. We will use this notation throughout the paper without redefining them.\\

\begin{theorem}
{\label {THM2}}
Assume a positive definite matrix ${\bf G}$ is $DP^k$ decomposable, that is ${\bf G = D - VV^T}$ as defined. Let ${\bf a}^*$ be a solution to  
$$\argmin_{{\bf a}\in \mathbb{Z}^n\backslash{\{0\}}} f({\bf a}) = {\bf a^TGa}$$
Then both the following statements are true:\\
{\bf a)} there exists a vector ${\bf x}\in \mathbb{R}^{k}$ such that  ${\bf a^*} - \frac{1}{2}\mathbf{1}< {\bf D^{-1}Vx} < {\bf a^*} + \frac{1}{2}\mathbf{1}$ and thus ${\bf a}^* = \round{{\bf D^{-1}Vx}}$. Or ${\bf a^*}$ must be a standard unit vector, up to a sign.\\
{\bf b)} $\|{\bf a}^*\| \le \psi = \sqrt{\frac{G_{min}}{\lambda_{min}}}$ where $G_{min}$ is the smallest diagonal element of ${\bf G}$ and $\lambda_{min}$ is the smallest eigenvalue of ${\bf G}$.
\end{theorem}

Note that \autoref{THM1} is a special case of \autoref{THM2} where $k=1$ , ${\bf D} = (1+P\|{\bf h}\|^2){\bf I}$ and ${\bf P} = P{\bf h h}^T$. The bound on the norm of ${\bf a}^*$ given by (\ref{eqn:CFbound}) is also equivalent to the bound given in part b) of \autoref{THM2} , since we have $\lambda_{min} = 1$ and $G_{min} \le {1 + P\|{\bf h}\|^2}$ in the lattice studied in \autoref{THM1}.

The Gram matrix in equation (\ref{eqn:CFMIMO}) also satisfies the constraints of \autoref{THM2}: Since ${\bf W}$ is a unitary matrix, ${\bf G}$ can be rewritten as ${\bf I} -  {\bf W(I-R)W}^T$. Clearly, ${\bf W(I-R)W}^T$ is of rank $k$ (since ${\bf I - R}$ has only $k$ non-zero diagonal entries), and positive semi-definite. The bound given by part b) of the theorem translates into $\|{\bf a}\| \le \sqrt{1 + P\gamma_{max}^2}$ where $\gamma_{max}$ is the maximum $\gamma_i$ value. This is because $G_{min}\le 1$ and the eigenvalues of ${\bf G}$ are easily seen to be equal to $\frac{1}{1 + P\gamma_i^2}$ (with the same eigenvectors as ${\bf HH}^T$) or 1.

\begin{rem}
{\label{Remark2}}
The SLV problem is easy to solve for diagonal Gram matrices: as the given basis is already orthogonal, the length of the shortest vector is the square root of the minimum diagonal element of the Gram matrix. \autoref{THM2} implies that subtracting a positive semi-definite low rank perturbation from a diagonal Gram matrix does not change the property of being polynomially solvable.
\end{rem}
\begin{rem}
{\label{Remark3}}
Deciding whether a matrix is $DP^k$ decomposable or not is outside of the scope of this work. Throughout this paper we will assume that the $DP^k$ decomposition of the Gram matrix is given a priori. The interested reader is referred to \cite{saunderson2012diagonal,shapiro1982weighted} for the state of the art algorithms which, under a set of conditions, can find the $DP^k$ decomposition of a matrix, with minimal $k$.
\end{rem}
\section{ALGORITHM II}
\label{sec:alg2}
For the case $k=1$ we presented an algorithm which finds precisely one point inside every interval in which $f$ is constant. For the general case, it is not clear to us how to find exactly one point per region. As a result we will present an algorithm which finds multiple points per region, while guaranteeing that first, every region has at least one representative point, and second, the number of points remains manageable, in the sense that it grows only as a polynomial function of $n$. \\
From \autoref{THM2} we know that the vector ${\bf a}^*$ satisfies the $2n$ inequalities:
$${\bf a^*} - \frac{1}{2}\mathbf{1}< {\bf D^{-1}Vx} < {\bf a^*} + \frac{1}{2}\mathbf{1}$$
for some ${\bf x}$. In other words, ${\bf x}$ belongs to the interior of the polytope described by these constraints. By analogy to the case $k=1$ , we start by finding the set of vertices of all such polytopes. Each vertex is the intersection of at least $k$ linearly independent hyperplanes of the form $c_{i} = ({\bf D^{-1}V})_{\{i\}}{\bf x}$, for half-integer $c_i$. Thus in order to find a vertex, we choose any set $ \pi\subseteq\{1,\dots ,n\}$ for which $|\pi|=k$ and $({\bf D^{-1}V})_{\pi}$ is full rank and solve ${\bf (D^{-1}V})_{\pi}{\bf x} = {\bf c}$ for ${\bf x}$ where the vector ${\bf c}$ consists of half integer elements. An arbitrary vertex ${\xi}_i$ thus falls in the following set:
\begin{equation}
\begin{split}
\xi_i \in \left\{ (({\bf D}^{-1}{\bf V})_\pi)^{-1}{\bf c} \;\middle| \;\pi\subseteq\{1,\dots,n\} \; ,\; |\pi| = k,\right. \\
\left. ({\bf D}^{-1}{\bf V})_\pi \text{ full rank} \; , \; {\bf c}-\frac{1}{2}\mathbf{1} \in\mathbb{Z}^k \right\}
\end{split}
  \label{eqn:12}
\end{equation}
According to part b) of \autoref{THM2} not all such vertices need to be checked, since: $\|{\bf a}_\pi^*\| \le \|{\bf a}^*\|\le\psi$. Thus like in the case $k=1$ we only need to check the vertices where  
$$-(\psi+\frac{1}{2})\mathbf{1}<{\bf (D^{-1}V})_{\pi}{\bf x}<(\psi+\frac{1}{2})\mathbf{1}$$
 and so 
 $$ -(\ceil{\psi}+\frac{1}{2})\mathbf{1} \le {\bf c} \le (\ceil{\psi}+\frac{1}{2})\mathbf{1}$$
 Now we can define the sets of all vertices of interest, ${\Phi}_\pi$ and their union $\Phi$ as 
 \begin{equation*}
\Phi_\pi =  \left\{ (({\bf D}^{-1}{\bf V})_\pi)^{-1}{\bf c} \;\middle|\;  |{\bf c}| \le (\ceil{\psi}+\frac{1}{2})\mathbf{1} \; , \; {\bf c}-\frac{1}{2}\mathbf{1} \in\mathbb{Z}^k \right\}
\end{equation*}
$$\pi\subseteq\{1\dots n\} \; ,\; |\pi| = k \;, ({\bf D}^{-1}{\bf V})_\pi \text{ full rank}$$
\begin{align*}
\Phi_\pi &= \emptyset  \;,\; \;\pi\subseteq\{1\dots n\} \; ,\; |\pi| = k \;, ({\bf D}^{-1}{\bf V})_\pi \text{ rank deficient}\\
\Phi &= \bigcup_{\substack{\pi\subseteq\{1\dots n\}\\|\pi| = k}}\Phi_\pi
\end{align*}  

In the next phase of the algorithm we use this set of vertices to find a set of interior points of polytopes of interest. It is not clear to us how to find exactly one point per polytope. The major difficulty is to identify which vertex belongs to which polytope. But for our main goal of showing a
polynomial bound on complexity, this is immaterial. \\
In order to find {\it at least} one point in the interior of each polytope, we then consider
all possible combinations of k + 1 vertices. Assuming they form a simplex in ${\mathbb R}^k$,
we can then find an interior point of this simplex by taking the average of the k + 1 vertices. Note that if the chosen vertices lie in a $k$-dimensional space, then they do not form a simplex. Nonetheless the algorithm can check the average of these points, even if the theorem does not consider it a potential minimizer.

Since any convex polytope can be decomposed into simplexes, an interior point of all the
polytopes must have been found in this process.
The last step is to check the value of $f$ over all these candidate points.
In line with the theorem, one also has to separately check all the standard unit vectors. \\
This is summarized in Algorithm \ref{Algorithm 2}.\\

\begin{algorithm}
\begin{algorithmic}[1]
\Statex {{\bf Input:} Gram matrix ${\bf G}$ and its $DP^k$ decomposition, ${\bf D}$ and  ${\bf V}$ matrices as defined}
\Statex {{\bf Output:} $\bf {a^*}$}
\Statex{}
\Statex {\bf \underline {Initialization:}}
\State ${\bf u}_i:=$ standard unit vector in the direction of $i$-th axis
\State $\lambda_{\min}:=$ minimum eigenvalue of ${\bf G}$
\State $G_{\min}:=$ minimum diagonal element of ${\bf G}$
\State $\psi := \sqrt{\frac{G_{\min}}{\lambda_{min}}}$
\State $\Phi=\emptyset$ 
\State $f({\bf a}):={\bf a^TGa}$
\State $f_{\min} = G_{\min}$
\State ${\bf a^* }= {\bf u}_{\argmin(diag({\bf G}))}$
\Statex{}
\Statex {\bf \underline {Phase 1:}}
\For{all $\pi \subseteq \{1,\dots ,n\}$, $|\pi| = k$, and ${\bf (D^{-1}V)_\pi}$ full rank}
  \For{all ${\bf c}$ , $|{\bf c} |\le (\ceil{\psi}+\frac{1}{2})\mathbf{1}$ and ${\bf c}-\frac{1}{2}\mathbf{1} \in\mathbb{Z}^k $}
\State calculate ${\bf x}=(({\bf D^{-1}V})_\pi)^{-1} {\bf c}_{\pi}$
\State Set $\Phi = \Phi \cup\{\bf x\}$
\EndFor
\EndFor

\Statex{}
\Statex {\bf \underline {Phase 2:}}
\For{all possible choices of $k+1$ points in $\Phi$}{
\State	calculate {\bf p} = average of the points
\State	calculate $\bf {b = D^{-1}Vp}$
\State	calculate ${\bf a} = \round{{\bf b}}$
	\If {$f({\bf a})<f_{\min}$ AND ${\bf a}$ is not the all zero vector}{
\State set ${\bf a^* = a}$
\State set $f_{\min}= f({\bf a})$
}
\EndIf
}
\EndFor
\State \Return {${\bf a^*}$}
\end{algorithmic}
\caption[caption]{Finding the optimal coefficient vector, the general case}
\label{Algorithm 2}
\end{algorithm}

\subsection{Complexity Analysis}
\label{sec:complexity}

The running time of the algorithm is clearly dominated by phase 2, where all possible $k+1$ combinations of the points found in phase 1 are checked as potential vertices of a simplex. First we count the number of points found in phase 1. This number is given by
\begin{equation}
\tag{**}
\begin{split}
\sum_{\substack{\pi\subset\{1\dots n\} \\  |\pi|=k}}&(2\ceil{\psi}+2)^k = {n \choose k}(2\ceil{\psi}+2)^k \\ 
\le \frac{n^k}{k!}&(2\ceil{\psi}+2)^k = \frac{\left(2n(\ceil{\psi}+1)\right)^k}{k!}
\end{split}
\label{eqn:**}
\end{equation}
The number of loops in phase 2 is the number of possible choices of $k+1$ points out of all points found in the phase 1. It can be upper bounded using equation (\ref{eqn:**}):  
$${\frac{\left(2n(\ceil{\psi}+1)\right)^k}{k!}\choose k+1}\le \frac{\left(2n(\ceil{\psi}+1)\right)^{k(k+1)}}{(k!)^{k+1}(k+1)!}$$
In order to find the complexity of the algorithm, we need to multiply this number of loops with the running time of each loop. Inside the loop, calculating the vector ${\bf b}$ can be done in $O(kn)$ and $f({\bf a})$ can also be calculated in $O(kn)$ operations. Thus the complexity of the algorithm is 
$$O\left(kn \frac{\left(2n(\ceil{\psi}+1)\right)^{k(k+1)}}{(k!)^{k+1}(k+1)!}\right) = O\left(\frac{n \left(2n(\ceil{\psi}+1)\right)^{k(k+1)}}{(k!)^{k+2}}\right)$$

Since this expression is a polynomial function of $\ceil{\psi} = \left\lceil\sqrt{\frac{G_{min}}{\lambda_{min}}}\right\rceil$, we conclude that as long as $\frac{G_{min}}{\lambda_{min}}$ is upper-bounded by a polynomial function of $n$ the complexity of the algorithm is polynomial in $n$. As we saw, in the case of MIMO Compute-and-Forward, we have $\psi = \sqrt{1+P\gamma_{max}^2}$. Also, we know that $\gamma_{max}^2\le trace({\bf HH}^T)\le nkH_{max}^2$, where $H_{max}$ is the maximum of all $|H_{ij}|$ values. Thus, if ${H_{max}}$ is bounded by a polynomial function of $n$, the algorithm will be of polynomial complexity. This is truly the case in all practical models, such as Rayleigh fading or any model with bounded channel coefficients.

\addtolength{\textheight}{-5.4cm}   
                                  
\section{PROOF OF THEOREM 2}
\label{sec:Proof1}
\subsection*{\bf part a)}

First note that we can rewrite $f({\bf a}) = {\bf a^TGa}$ as follows:
$$f({\bf a}) = \sum_{i=1}^n (D_{ii}-P_{ii})a_i^{2} - 2\sum_{i=1}^n\sum_{j=1}^{i-1}P_{ij}a_ia_j $$
Assume that we already know the optimal value for all $a_i^*$ elements except for one element, $a_j$. Note that $f$ is a convex parabola in $a_j$ (this is because $D_{jj}-P_{jj} = G_{jj}$ is a diagonal element of a positive definite matrix) thus the optimal integer value for $a_j$ is the closest integer to its optimal real value. By taking partial derivative with respect to $a_j$, the optimal real value of $a_j$ is easily seen to be equal to 
$$\frac{\sum_{i=1, i\neq j}^nP_{ij}a_i^*}{D_{jj} - P_{jj}}.$$
Taking the closest integer to the real valued solution, we find:
\begin{equation}
\tag{I}
\Rightarrow a_j^*=\left\lceil\frac{\sum_{i=1, i\neq j}^nP_{ij}a_i^*}{D_{jj} - P_{jj}}\right\rfloor \;\; OR \;\;a_j^*=\left\lfloor\frac{\sum_{i=1, i\neq j}^nP_{ij}a_i^*}{D_{jj} - P_{jj}}\right\rceil
\label{eqn:I}
\end{equation}
Due to the symmetry of the parabola, both functions return equally correct solutions for $a_j^*$. \\
Note that this expression must be true for any $j$: If for ${\bf a^*}$ and for some $j$, $a_j^*$ does not satisfy at least one of these two equations, we can achieve a strictly smaller value over $f$ by replacing $a_j^*$ with the value given above, and so ${\bf a}^*$ cannot be optimal. The only situation where this logic fails is when in the optimal vector we have: $a_i^* = 0$ , $i=1\dots n$ , $i\neq j$. In this case, replacing the value of $a_j^*$ with its round expression will result in the all zero vector, ${\bf a}^*= {\bf 0}$. Hence, the case where ${\bf a}^*$ is zero except in one element requires separate attention, as pointed out by the theorem. Under this assumption, $f({\bf a}^*) = G_{jj}a_j^{*2}$. Thus it must be that ${|a_j^*| } = 1$, and so ${\bf a}^*$ is a standard unit vector, up to a sign.\\
Returning to the general case of ${\bf a}^*$ and from (\ref{eqn:I}) we have that:
\begin{align*}
\tag{II}
a_j^*+\frac{1}{2}&\ge \frac{\sum_{i=1, i\neq j}^nP_{ij}a_i^*}{D_{jj} - P_{jj}} \;\;\;,\;\;and \\
\tag{III}
a_j^*-\frac{1}{2}&\le \frac{\sum_{i=1, i\neq j}^nP_{ij}a_i^*}{D_{jj} - P_{jj}}
\end{align*}
Starting with equation (II), we multiply both sides by the denominator, and add the term $a_j^*P_{jj}$ to obtain:
\begin{align*}
(a_j^*+\frac{1}{2})D_{jj}&\ge \sum_{i=1}^nP_{ij}a_i^*+\frac{1}{2}P_{jj}\\
\end{align*}
Dropping the non-negative term $\frac{1}{2}P_{jj}$ we conclude 
$$(a_j^*+\frac{1}{2})D_{jj}\ge \sum_{i=1}^nP_{ij}a_i^*$$
Now we show that this inequality is strict, even if $P_{jj} = 0$. Due to the fact that ${\bf P}$ is positive semi-definite, we must have that if $P_{jj} = 0$ then $P_{ij} = 0$ , $i=1\dots n$. Thus in that case, the inequality turns into $(a_j^*+\frac{1}{2})D_{jj}\ge 0$. But we have that $a_j^*$ is an integer and $D_{jj}-P_{jj}>0$ thus $D_{jj}>0$. So, $(a_j^*+\frac{1}{2})D_{jj}$ cannot be equal to zero and this inequality must be strict. As a result, we have:
\begin{align*}
(a_j^*+\frac{1}{2})D_{jj}&> \sum_{i=1}^nP_{ij}a_i^*\\
\Rightarrow (a_j^*+\frac{1}{2})&> \frac{\sum_{i=1}^nP_{ij}a_i^*}{D_{jj}} \; , \; j=1\dots n
\end{align*}
Writing this inequality in vector format, we obtain
\begin{equation}
\tag{IV}
{\bf a}^*+\frac{1}{2}\mathbf{1}> {\bf D}^{-1}{\bf P}^T{\bf a}^* = {\bf D}^{-1}{\bf V}({\bf V}^T{\bf a}^*)
\label{eqn:IV}
\end{equation}
In a similar fashion one can show that equation (III) results in
\begin{equation}
\tag{V}
\Rightarrow {\bf a}^*-\frac{1}{2}\mathbf{1}< {\bf D}^{-1}{\bf P}^T{\bf a}^* = {\bf D}^{-1}{\bf V}({\bf V}^T{\bf a}^*)
\label{eqn:V}
\end{equation}
Defining ${\bf x} = {\bf V}^T{\bf a}^*$, it follows from (\ref{eqn:IV}) and (\ref{eqn:V}) that
$${\bf a^*} - \frac{1}{2}\mathbf{1}< {\bf D^{-1}Vx} < {\bf a^*} + \frac{1}{2}\mathbf{1}$$
$$\Rightarrow {\bf a}^* = \round{{\bf D^{-1}Vx}}$$
This completes the proof of part a).

\subsection*{\bf part b)}
First note that
$$f({\bf a^*}) = {\bf a^*}^T{\bf G a^*} \ge \lambda_{min}\|{\bf a}^*\|^2$$
By simply choosing ${\bf a}$ to be the $i$-th standard unit vector, we have $f({\bf a}) = G_{ii}$. Thus:
$$G_{min}\ge f({\bf a}^*) \ge \lambda_{min}\|{\bf a^*}\|^2 $$
from which we can conclude
\begin{align*}
\|{\bf a^*}\| & \le\sqrt{\frac{{G}_{min}}{\lambda_{min}}}
\end{align*}
which is the claim made by part b) of the theorem.

\section{CONCLUSIONS AND FUTURE WORK}
\label{sec:conclusion}
In this paper we introduced an exact algorithm of polynomial complexity for solving the special case of SLV problem which appears in the context of Compute-and-Forward. We then generalized our results to the case of MIMO Compute-and-Forward. There are several possibilities to continue this work. The results may be extendable to more general lattices. Furthermore such Gram matrices may be used as a point of reference for approximating the shortest vector in a wider range of lattices. Finally, we conjecture that particular choices of the matrix ${\bf V}$ in decomposition of the Gram matrix may allow for a more efficient algorithm by establishing simple relations between the $x_i$ values.

\section*{ACKNOWLEDGMENT}

We would like to thank Chien-Yi Wang for his interesting ideas which helped us in different stages of this work. We would also like to thank him, Robby McKilliam and Cheng Wang for their help with reviewing the paper. This work has been supported in part by the European Union under ERC
Starting Grant 259530-ComCom.





\bibliographystyle{IEEEtran}
\bibliography{IEEEfull,root}

\end{document}